\documentclass{article}

\usepackage{blindtext} 

\usepackage[T1]{fontenc} 
\usepackage{microtype} 

\usepackage[english]{babel} 

\usepackage[hmarginratio=1:1,top=32mm,columnsep=20pt]{geometry} 
\usepackage[hang, small,labelfont=bf,up,textfont=it,up]{caption} 
\usepackage{booktabs} 

\usepackage{lettrine} 
\usepackage{amsmath}
\usepackage{float} 
\usepackage{enumitem} 
\setlist[itemize]{noitemsep} 

\usepackage{abstract} 

\usepackage{titlesec} 
\renewcommand\thesection{\Roman{section}} 
\renewcommand\thesubsection{\roman{subsection}} 
\titleformat{\section}[block]{\large\scshape\centering}{\thesection.}{1em}{} 
\titleformat{\subsection}[block]{\large}{\thesubsection.}{1em}{} 

\usepackage{fancyhdr} 
\usepackage{scalerel,stackengine}
\usepackage{titling} 

\usepackage{hyperref} 

\pretitle{\begin{center}\Huge\bfseries} 
\posttitle{\end{center}} 
\title{Speaker Clustering With Neural Networks And Audio Processing} 
\author{%
\textsc{Maxime Jumelle, Taqiyeddine Sakmeche} \\[1ex] 
\normalsize AIPCloud - \url{http://aipcloud.io} \\ 
  \normalsize \{\href{mailto:maxime@aipcloud.io}{maxime}, \href{mailto:taqiyeddine@aipcloud.io}{taqiyeddine}\}@aipcloud.io
}
\date{November 2017} 
\renewcommand{\maketitlehookd}{%
\begin{abstract}
\noindent Speaker clustering is the task of differentiating speakers in a recording. In a way, the aim is to answer "who spoke when" in audio recordings. A common method used in industry is feature extraction directly from the recording thanks to MFCC features, and by using well-known techniques such as Gaussian Mixture Models (GMM) and Hidden Markov Models (HMM). In this paper, we studied neural networks (especially CNN) followed by clustering and audio processing in the quest to reach similar accuracy to state-of-the-art methods.
\end{abstract}
}

\begin{document}

\maketitle


\section{Introduction}

Convolutional Neural Networks \cite{LeCun:1989:BAH:1351079.1351090} (CNN) are very successful when it comes to classification tasks in high-dimensional spaces, especially for image classification. Every now and then new ways to increase their efficiency are found. They are very useful to show hidden features from the data they are fed. As a result, and because CNN can also be used to achieve more advanced tasks with feature representation as in \cite{DBLP:journals/corr/AthiwaratkunK15}, we created a model to cluster unknown speakers in which CNN plays the role of the feature extractor. It is fed the raw spectrogram, instead of MFCC (Mel Frequency Cepstral Coefficients) \cite{MFCC_article} which is usually used for speech recognition tasks.
\vspace{0.3cm}
\\
Before the avent of neural networks, common methods involved Gaussian Mixture Models (GMM) \cite{speaker_diarization_GMM} and other derivative methods such as Hidden Markov Models (HMM) \cite{kl_hmm} that produced best results. In this paper, we try to create a CNN that can extract useful features in voice recognition. We will then train a classic CNN on the TIMIT corpus dataset \cite{timit}, followed by a fully-connected layers (Multi Layer Perceptrons) in order to classify. The objective is to make non-linear data become linearly separable. We will then use the CNN as a feature extractor, and retrieve the output of specific layers which will be used as feature vectors. With those feature vectors, we will finally use clustering algorithms to cluster each vector and then create groups corresponding to each speaker.

\section{Methods}

\subsection{Audio pre-processing}

Usually input data that was fed into models was processed thanks to MFCC features in the field of speech recognition as in \cite{DBLP:journals/corr/abs-1003-4083} or \cite{online_speaker_diarization}. Recent papers demonstrates that we can achieve comparables results with CNN and that we can obtain linearly separable data thanks to them, whether using raw signal data \cite{DBLP:journals/corr/PalazMC14} as input or the mel-spectrogram \cite{speaker_identification_and_clustering}.

\begin{figure}[!htbp]
\centering
\includegraphics[scale=0.25]{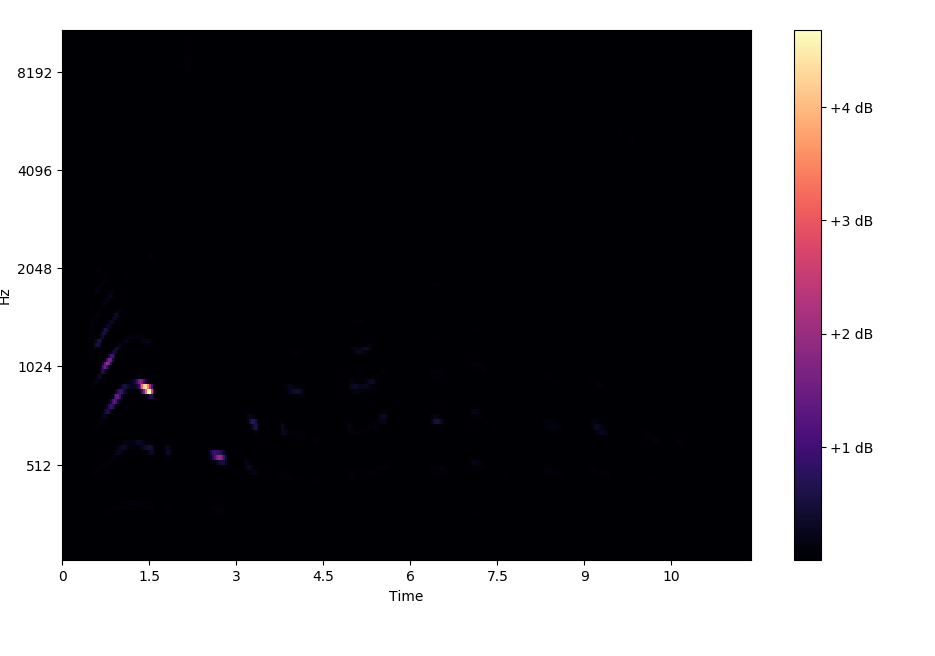}
\includegraphics[scale=0.25]{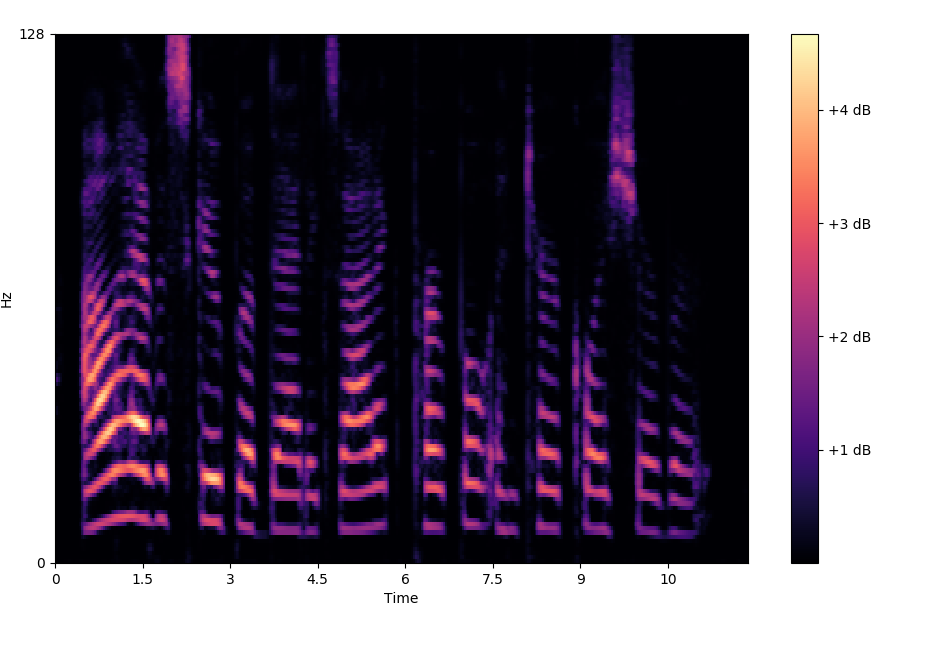}
\caption{On the left, the spectrogram of a sample from the TIMIT corpus dataset, on the right, the mel-spectrogram. We can see that mel-spectrogram shows interesting voice features instead of the spectrogram.}
\label{logspectrogram}
\end{figure}

\subsection{Feature extraction}

In this model, instead of considering a vector of MFCC coefficients as input, we directly use the mel-spectrogram such as a classical CNN would classify the 2D-matrix as an image. Here, we only have one channel, corresponding to the decibels : inputs of the CNN are arrays of a certain dimension. We choose to study packs of frames of length 1 second, with a 10 ms step window, that is to say we expect $(128, 100)$ as input arrays. The network will be trained on the TIMIT dataset. As shown in Figure \ref{logspectrogram}, samples from dataset are not significant enough to be fed in the network without audio processing. To overcome this situation, we compute the log-spectrogram with $f(x)=\log (1+10,000x)$, which allows us to efficiently extract voice characteristics of each speaker.

\section{Model and experiments}

\subsection{CNN Architecture}

We have a training set $\mathbf{X}=(\mathbf{X_1}, \dots, \mathbf{X_n})$ of $n$ samples where $\mathbf{X_1}$ is a matrix of dimensions $128 \times 100$ corresponding to snippets of $1$ seconds. We also define the label set $Y=(y_1, \dots, y_n)$ where $y_i \in \{0, \dots, 549\}, 1 \leq i  \leq n$ denotes the speaker identifier. We first need to train the CNN as a classification problem, so we would like to be able to find a classifier $g$ such that $g(\mathbf{X}) \approx Y$. Since it is almost impossible to obtain the true classifier $g$, we will construct, thanks to the CNN, an approximation $g^*$ of it, trying to keep the same idea that $g^*(\mathbf{X}) \approx Y$. \\
As we mentioned earlier, our goal is to train a CNN as a feature extractor. We split the TIMIT dataset into a training set of $550$ speakers and a validation set of the other speakers.
\begin{figure}[H]
\centering
\includegraphics[scale=0.37]{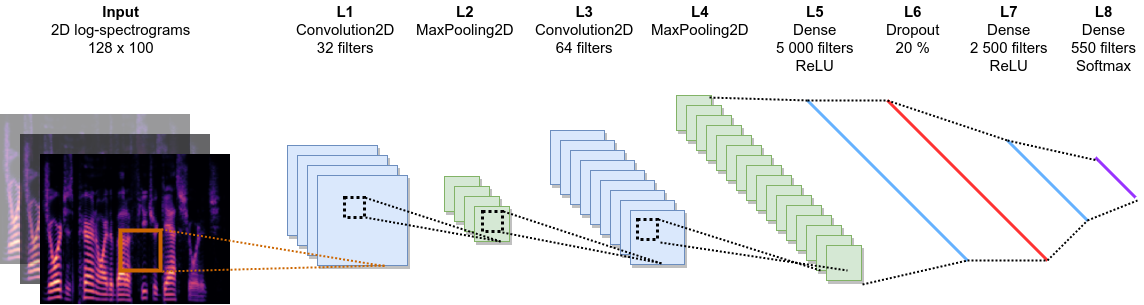}
\caption{Architecture of the CNN.}
\label{CNN}
\end{figure}
First, we use a two stage convolution filter in order to obtain features directly from the log-spectrograms. Each stage involves a 2D convolution layer followed by a 2D max-pooling. For the first stage, we use $32$ filters on the convolution layer with a kernel of size $L_F=3$. The max-pooling layer has a pool-size of $L_P=4$ pixels. On the second stage, we only double the filters using $64$ filters as parameters for kernel and pool-size are the same. Let $F$ be the filter matrix of dimension $L_F \times L_F$ in a convolutional layer. We choose to use a linear activation function. If we denote $o_{i,j}$ the $i$-th row and $j$-th column of the output of the first convolutional layer
$$o_{i,j}=\sum_{k,l=1}^{L_F} \mathbf{X_1}^{(i+k-1,j+l-1)} F_{i, j}$$
We then proceed with a max-pooling layer, and the output of this layer $m_{i, j}$ is given by
$$m_{i,j}=\max_{i-\lfloor \frac{L_P}{2}\rfloor\leq k \leq i+\lfloor \frac{L_P}{2}\rfloor} \quad \max_{j-\lfloor \frac{L_P}{2}\rfloor \leq l \leq j+\lfloor \frac{L_P}{2}\rfloor} o_{k,l}$$
Right after the first dense layer, composed of $5000$ filters with ReLU as an activation function (L5), we chose to add a Dropout layer which is useful to prevent from overfitting with a rate of $\eta=0.2$. Finally, we add the last dense layer, composed of $2500$ filters also with ReLU as an activation function (L7), which will be the output layer of the CNN as a feature extractor. Since we would like to train the CNN on a classification task, we end the model with a softmax layer (L8) of $550$ filters, which corresponds to the number of speakers. The model is designed in Figure~\ref{CNN}. \\
During the training phase, we considered a Stochastic Gradient Descent optimizer with categorical cross-entropy as a loss function.

\begin{figure}[H]
\centering
\includegraphics[scale=0.5]{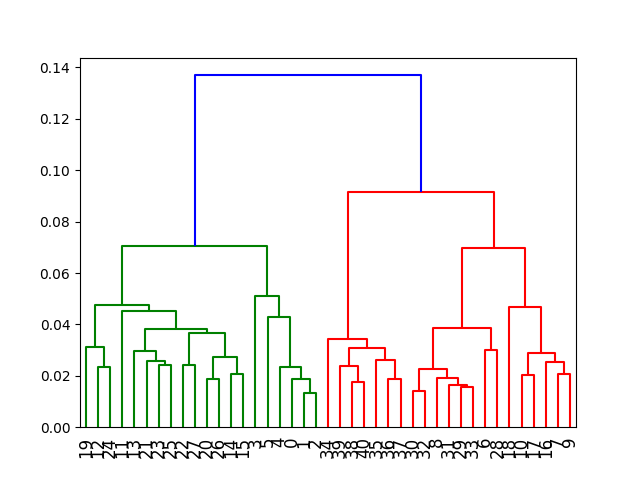}
\caption{The dendrogram of the different clusters available.}
\label{Dendrogram}
\end{figure}

\subsection{Manifold learning and clustering}

Our major idea was to suppose that thanks to the CNN as a feature extractor, one second long snippets data points from the unknown input data, are now linearly separable. As a result, we will obtain filter outputs from the last fully-connected layer (L7), and in the process get a vector in the feature space $\mathcal{X}$.
\begin{figure}[H] 
\centering
\includegraphics[scale=0.4]{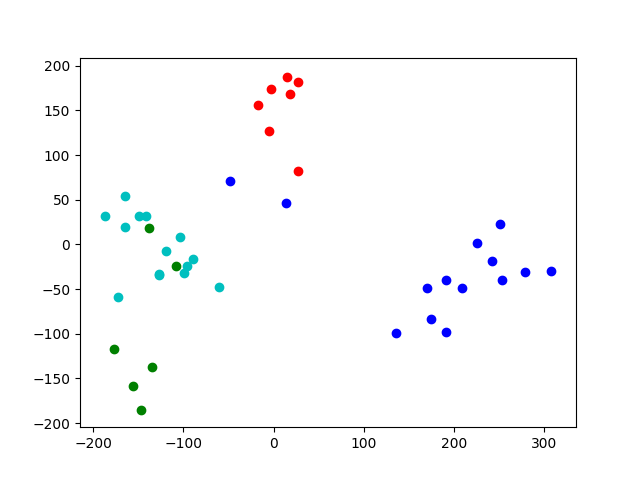}
\includegraphics[scale=0.4]{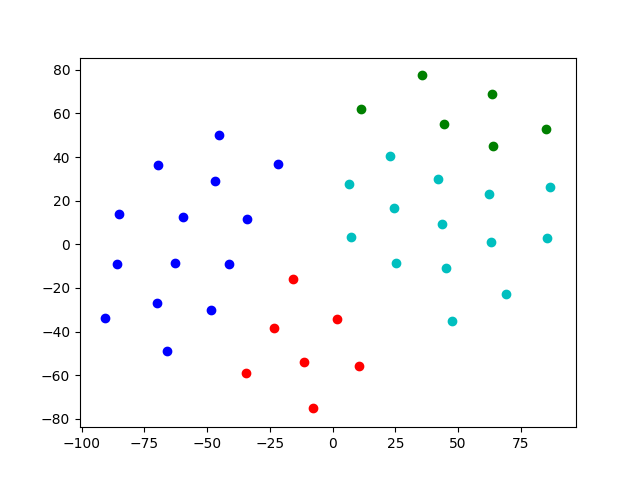}
\caption{Dimensionality reduction of data points. The left graph is the PCA reduced space and the right graph is the t-SNE reduced space with the cosine metric. The fact that groups are more easily visualizable on the right image shows the non-linearity aspect of the data. Each color corresponds to a different speakers.}
\label{Dim_reduction}
\end{figure}
\vspace{0.3cm}

In this context, it can be useful to use t-SNE \cite{vanDerMaaten2008} in order to easily identify different clusters. In Figure \ref{Dim_reduction}, we compared dimensionality reduction with PCA and t-SNE. Although we can detect some clusters with PCA, t-SNE's ability to detect clusters in high-dimensional space gives more satisfying results. The t-SNE representation comforts the idea that CNN architecture produces linearly separable data.
\\
On the feature space, we use a hierarchical clustering with the cosine metric \cite{hierarchical_clustering} which produces a
dendrogram. Thanks to this, we can detect how many speakers are in the file splitting at the good
branch. We then have to find the minimal gap between two consecutives groups to obtain the best
estimation of clusters.

\subsection{Speaker intermediate frames}

Thanks to the clustering from the previous state, we are now able to identify which speaker is talking at every second. Needless to say that one second is not accurate enough. Since we only want to know the moment when a speaker A stops talking and another speaker B starts talking, it is more interesting to take a look at voice characteristics such as pitch class. To perform this segmentation, we start to produce 2 seconds frame length centered around the second when speakers A stops talking and speaker B begins. We compute the Chroma feature \cite{MuellerEwert11_ChromaToolbox_ISMIR} on these frames that gives pitch classes.
\begin{figure}[H]
\centering
\includegraphics[scale=0.4]{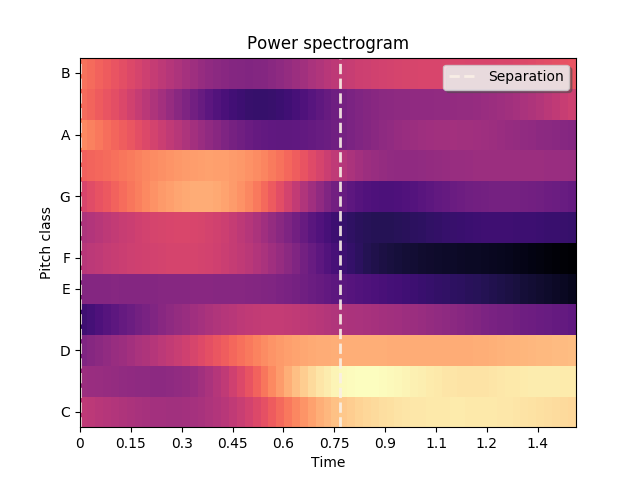}
\caption{Chroma plot of a 2 seconds frame.}
\label{Chroma}
\end{figure}
From this, we perform a bottom-up temporal segmentation to get the time of separation between speakers A and B.

\subsection{Speaker identification}

\begin{figure}[!htbp]
\centering
\includegraphics[scale=0.5]{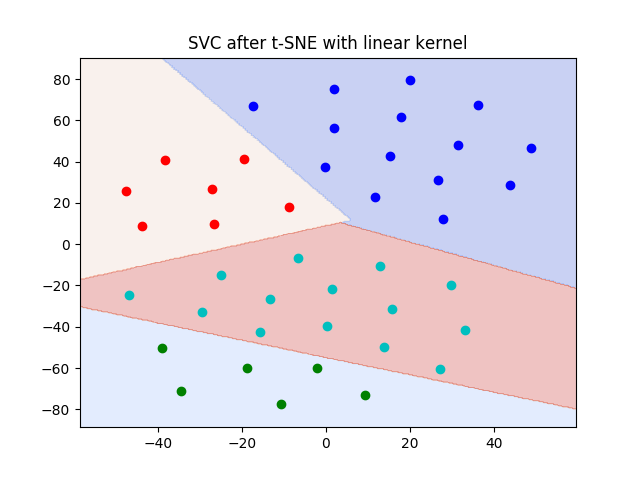}
\caption{Example of a Support Vector Clustering on t-SNE reduced space with linear kernel.}
\label{SVC}
\end{figure}

Once we applied a clustering algorithm on our feature space, we are now able to train a classifier such as an SVM in order to identify speaker from a few seconds recording on each of them. We then use a Support Vector Clustering \cite{svc} with a linear kernel using One-vs-One approach. As shown in Figure \ref{SVC}, there is a good separations between clusters and the identification task can be easily performed with a trained SVC.

\section{Conclusion}

Although we obtain high accuracy both on training and test sets, real-life applications point out the fact that our model encounters difficulties when it comes to analyze voices in presence of noise or resonance. Moreover, because of our audio pre-processing, our model is unable to detect short interventions (such as 1 or 2 seconds). We can see two main improvements :
\begin{itemize}
\item We could fine-tune network's parameters to set optimal hyper-parameters to strengthen analysis on short frames.
\item Adjust the current model so that 1 or 2 seconds length frame could be computed efficiently.
\end{itemize}
We are also looking at the freshly developped Capsule Network \cite{DBLP:journals/corr/abs-1710-09829} of Geoffrey Hinton and his team, trying to improve accuracy.


\bibliographystyle{unsrt}
\bibliography{main}

\begin{thebibliography}{10}

\bibitem{LeCun:1989:BAH:1351079.1351090}
Y.~LeCun, B.~Boser, J.~S. Denker, D.~Henderson, R.~E. Howard, W.~Hubbard, and
  L.~D. Jackel.
\newblock Backpropagation applied to handwritten zip code recognition.
\newblock {\em Neural Comput.}, 1(4):541--551, December 1989.

\bibitem{DBLP:journals/corr/AthiwaratkunK15}
Ben Athiwaratkun and Keegan Kang.
\newblock Feature representation in convolutional neural networks.
\newblock {\em CoRR}, abs/1507.02313, 2015.

\bibitem{MFCC_article}
Sayf Majeed, HAFIZAH HUSAIN, SALINA ABDUL~SAMAD, and Tarik Idbeaa.
\newblock Mel frequency cepstral coefficients (mfcc) feature extraction
  enhancement in the application of speech recognition: A comparison study.
\newblock 79:38--56, 09 2015.

\bibitem{speaker_diarization_GMM}
Michael Betser, Frédéric Bimbot, Mathieu Ben, and Guillaume Gravier.
\newblock Speaker diarization using bottom-up clustering based on a
  parameter-derived distance between adapted gmms.
\newblock 01 2004.

\bibitem{kl_hmm}
S.~Madikeri and H.~Bourlard.
\newblock Kl-hmm based speaker diarization system for meetings.
\newblock pages 4435--4439, April 2015.

\bibitem{timit}
J.~S. Garofolo, L.~F. Lamel, W.~M. Fisher, J.~G. Fiscus, D.~S. Pallett, and
  N.~L. Dahlgren.
\newblock {DARPA} {TIMIT} acoustic phonetic continuous speech corpus {CDROM},
  1993.

\bibitem{DBLP:journals/corr/abs-1003-4083}
Lindasalwa Muda, Mumtaj Begam, and I.~Elamvazuthi.
\newblock Voice recognition algorithms using mel frequency cepstral coefficient
  {(MFCC)} and dynamic time warping {(DTW)} techniques.
\newblock {\em CoRR}, abs/1003.4083, 2010.

\bibitem{online_speaker_diarization}
Giovanni Soldi, Christophe Beaugeant, and Nicholas Evans.
\newblock Adaptive and online speaker diarization for meeting data.
\newblock pages 2112--2116, 08 2015.

\bibitem{DBLP:journals/corr/PalazMC14}
Dimitri Palaz, Mathew Magimai{-}Doss, and Ronan Collobert.
\newblock Learning linearly separable features for speech recognition using
  convolutional neural networks.
\newblock {\em CoRR}, abs/1412.7110, 2014.

\bibitem{speaker_identification_and_clustering}
Y.~Lukic, C.~Vogt, O.~Dürr, and T.~Stadelmann.
\newblock Speaker identification and clustering using convolutional neural
  networks.
\newblock pages 1--6, Sept 2016.

\bibitem{vanDerMaaten2008}
Laurens van~der Maaten and Geoffrey Hinton.
\newblock Visualizing data using {t-SNE}.
\newblock {\em Journal of Machine Learning Research}, 9:2579--2605, 2008.

\bibitem{hierarchical_clustering}
Sébastien Marcel.
\newblock “hierarchical speaker clustering methods for the nist i-vector
  challenge”, elie khoury, laurent el shafey, marc ferras and sebastien
  marcel, odyssey: The speaker and language recognition workshop.
\newblock 01 2014.

\bibitem{MuellerEwert11_ChromaToolbox_ISMIR}
Meinard M{\"u}ller and Sebastian Ewert.
\newblock {C}hroma {T}oolbox: {MATLAB} implementations for extracting variants
  of chroma-based audio features.
\newblock In {\em Proceedings of the 12th International Conference on Music
  Information Retrieval ({ISMIR})}, Miami, USA, 2011, to appear.

\bibitem{svc}
Asa Ben-Hur, David Horn, Hava Siegelmann, and Vladimir Vapnik.
\newblock Support vector clustering.
\newblock 2:125--137, 11 2001.

\bibitem{DBLP:journals/corr/abs-1710-09829}
Sara Sabour, Nicholas Frosst, and Geoffrey~E. Hinton.
\newblock Dynamic routing between capsules.
\newblock {\em CoRR}, abs/1710.09829, 2017.

\end{thebibliography}

\end{document}